\documentclass[aps,prb,twocolumn,groupedaddress,amssymb,superscriptaddress]{revtex4}

\usepackage[dvips]{graphicx}
\usepackage{bm}      
\usepackage[normalem]{ulem}    
\newcommand{\ybcoy}{Y\-Ba$_2$\-Cu$_3$\-O$_{6+y}$}
\newcommand{\ybco}{Y\-Ba$_2$\-Cu$_3$\-O$_{6}$}
\newcommand{\calabalacuo}{Ca$_{x}$La$_{1.25}$Ba$_{1.75-x}$Cu$_3$O$_{6+y}$}
\newcommand{\lscozn}{La$_{2-x}$Sr$_x$\-Cu$_{1-z}$Zn$_z$O$_4$}
\newcommand{\ycabcoy}{Y$_{1-x}$Ca$_x$\-Ba$_2$\-Cu$_3$\-O$_{6+y}$}
\newcommand{\ycabcofive}{Y$_{0.95}$Ca$_{0.05}$\-Ba$_2$\-Cu$_3$\-O$_{6+y}$}

\newcommand{\yeubco}{Y$_{0.92}$Eu$_{0.08}$\-Ba$_2$\-Cu$_3$\-O$_{6+y}$}
\newcommand{\yndbco}{Y$_{0.925}$Nd$_{0.075}$\-Ba$_2$\-Cu$_3$\-O$_{6+y}$}
\newcommand{\ycabco}{Y$_{1-x}$Ca$_y$\-Ba$_2$\-Cu$_3$\-O$_{6}$}
\newcommand{\lsco}{La$_{2-x}$Sr$_x$\-CuO$_4$}
\newcommand{\lnsco}{La$_{1.6-x}$\-Nd$_{0.4}$\-Sr$_x$\-CuO$_4$}
\newcommand{\lco}{La$_{2}$\-CuO$_4$}
\newcommand{\nsm}{NL$\sigma$M}
\newcommand{\msr}{$\mu$SR}
\newcommand{\old}[1]{ } 

\begin{document}


\title{The magnetic states of lightly hole-doped cuprates in the clean limit. }


\author{F.~Coneri}
\affiliation{Dipartimento di Fisica, Universit\`a di Parma, Viale G.Usberti, 7A, I-43100 Parma, Italy}
\author{S.~Sanna}
\email[]{Samuele.Sanna@unipv.it}
\affiliation{Dipartimento di Fisica A.~Volta, Universit\`a di Pavia, Via Bassi, 6, I-27100 Pavia, Italy}
\affiliation{Dipartimento di Fisica, Universit\`a di Parma, Viale G.Usberti, 7A, I-43100 Parma, Italy}
\author{K.~Zheng}
\affiliation{Dipartimento di Fisica, Universit\`a di Parma, Viale G.Usberti, 7A, I-43100 Parma, Italy}
\author{J.~Lord}
\affiliation{ISIS Facility, STFC-Rutherford Appleton Lab, HSIC, Didcot, OX110QX, U. K.}
\author{R.~De Renzi}
\affiliation{Dipartimento di Fisica, Universit\`a di Parma, Viale G.Usberti, 7A, I-43100 Parma, Italy}



\date{\today}

\begin{abstract}
We have performed extensive zero field $\mu$SR experiments on pure \ybcoy\ and diluted Y-rare-earth substituted \yeubco\ and \yndbco\ at light hole-doping.  A common magnetic behavior is detected for all the three families, demonstrating negligible effects of the isovalent Y-substituent disorder. Two distinct regimes are identified, separated by a crossover, whose origin is attributed to the concurrent thermal activation of spin and charge degrees of freedom: a thermally activated and a re-entrant antiferromagnetic regime. The peculiar temperature and hole density dependence of the
magnetic moment $m(h,T)$ fit a model with a (spin) activation energy
for the crossover between the two regimes throughout the entire investigated range. The magnetic moment is suppressed by a simple
dilution mechanism both in the re-entrant regime ($0\le h \le 0.056$) and in the so-called Cluster Spin Glass state coexisting with superconductivity ($0.056<h\lesssim 0.08$). We argue a common magnetic ground state for these two doping regions and dub it {\em frozen} antiferromagnet.
Conversely either frustration or finite-size effects prevail in the {\em thermally activated} antiferromagnetic state, that vanishes at the same concentration where superconductivity emerges, suggesting the presence of a quantum critical point at $h_c=0.056(2)$.

\end{abstract}


\maketitle


\section{\label{sec:intro}Introduction}
The magnetic ground state of cuprates at very low doping is qualitatively understood as that of a frustrated\cite{AharonyPRL1988} two dimensional Heisenberg antiferromagnet (2DHAF). The 2DHAF nature was recognized very early\cite{ChakravartyPRL1988} from the successful description that its long wavelength approximation, the nonlinear $\sigma$ model (\nsm), provides of the finite correlation lengths detected by neutron scattering\cite{KeimerPRB1992} above the transition to three dimensional (3D) order.

The wave function of a hole in the N\'eel AF background leads to a local spin singlet\cite{Zhang-RicePRB1988}, i.e. it neutralizes a Cu spin, but its dynamical effects reduce the ordering temperature $T_N$ rather more effectively than a mere magnetic-site dilution. The abrupt reduction of the ordering temperature $T_N$ (as detected very early on by implanted muons\cite{BrewerPRL1988}) corresponds also to a drop of the static magnetic moment\cite{EndohPRB1988,Rossat-MignodPB1990,TranquadaPRB1989}
at $Q=\left(\frac 1 2\, \frac 1 2\right)$. The strong frustrating effect of holes in the 2DHAF background may be understood in several ways. Staggered spin spirals\cite{LuescherPRL2007} have been predicted and their hole density dependence reproduces the incommensurate magnetic scattering\cite{MatsudaPRB2002} detected in \lsco. Stripes have also been predicted, either as charged spin solitons\cite{ZaanenPRB1989} or from electronically driven phase separation.\cite{EmeryPRL1990} They have been detected\cite{TranquadaNature1995} in  specific doping and distortive conditions (e.g. \lnsco), and are possibly present also as correlated fluctuations.

Notwithstanding this partially quantitative agreement between models and experiments we are still far from a complete understanding of the low-doping  Mott Hubbard insulator and the number of open questions grows on approaching and then crossing the metal-insulator (MI) transition. They concern the precise nature of the magnetic ground state versus temperature, the origin of the spin glass phase intervening at slightly higher doping, the presence of static magnetic moments in the very low-doping superconducting state, and ultimately the competitive or cooperative role of magnetic interactions in the strongly correlated superconductor.

The theoretical difficulties increase with hole density, but the scarcity of experimental studies in the less puzzling insulating range prevents precise checks across the whole insulating phase diagram. In addition the real cuprate compounds, as opposed to the idealized CuO$_2$ model layer, are already distinguished at very low doping by structure and intrinsic disorder.\cite{NiedermayerPRL1998} It is now clear,\cite{AndoPRL2001,SannaPRL2004,DagottoNature2005,AlvarezPRB2005,AlloulRMP2009} although not always recognized, that \ybcoy, where the charge reservoir is farther removed from the CuO$_2$ layer, represents the closest physical system to the clean limit, whereas randomly placed  divalent cations make \lsco\ and \ycabco\ dirty limit cases, where disorder, increasing with doping, actively modifies the scenario.

In this work we decided to investigate extensively the low-doping range, starting from the clean limit case. Besides pure \ybcoy\ (Y100\% in the following) we addressed two additional compositions, to investigate the effect of small lattice distortion on the magnetic phase diagram by substitution of Y with a rare earth: \yeubco\ (Eu8\%), yielding  a very small mismatch of the cation radii, and \yndbco\ (Nd7.5\%), yielding equivalent mismatch to that of the 5\% Ca substitution investigated in a parallel work. \cite{SannaArXiv2009} It turns out that isovalent substitution affects  negligibly both the magnetic behavior and the MI transition.

Muon spin spectroscopy (\msr) was selected as a local probe sensitive to static magnetic order, even with small moments, and to local spin rearrangements. This technique reveals a number of evident trends, yielding quantitative constraints for a model of doped 2DHAF appropriate for clean cuprates. We carefully identify two different magnetic regimes and measure their distinctive parameters, evidencing their critical behavior with doping in the whole investigated range.

The paper is organized in three main sections: Sec.~\ref{sec:2DHAF} briefly reviews models and experimental facts; Sec.~\ref{sec:results} describes the muon results in terms of a model of thermally activated charge excitations; Sec.~\ref{sec:discussion} discusses the implication of the data and of the model on the magnetic state in doped Mott-Hubbard insulating cuprates. The appendices describe the relation between the magnetic field at the muon site and the average staggered Cu moment, they provide details on the samples and on the muon data analysis.

\section{\label{sec:2DHAF} The doped $S=\frac 1 2$ 2D Heisenberg antiferromagnet.}

Copper perovskites, characterized by one or more CuO$_2$ layers of square plaquettes,  have a common starting point in the two dimensional (2D) N\'eel order of their parent compound, technically a half-filled Mott-Hubbard insulator of the charge transfer \cite{OgataRPP2008} type, dominated by a large isotropic antiferromagnetic (AF) exchange $J<0$ along the in-plane bonds, with a much reduced coupling $J^\prime$ along the $c$ axis.

The predicted transition for a cubic system with the same $J$ along all the three directions (3D) would be rather large, $T_{N0}\approx 1500$ K, whereas the observed transitions are $T_N=310,420$ K for \lco\ and \ybco, respectively, and very large in-plane correlation lengths\cite{EndohPRB1988} persist far above $T_N$. This behavior, previously encountered in localized spin systems, such as the fluorite perovskites,\cite{BirgeneauPRB1970} is described at high temperature by the 2D Heisenberg (2DHAF) model. For a pure 2D system the Mermin-Wagner theorem \cite{Mermin-WagnerPRL1966} predicts $T_N=0$, but three-dimensional order sets in at an intermediate temperature, scaling roughly as $1/\log(J/J^\prime)$.  The ideal 2DHAF behavior with its large quantum spin reduction for $S=1/2$ and the high temperature correlations are quantitatively described\cite{ChakravartyPRL1988} in the \nsm\ long-wavelength approximation.  The spin stiffness $\rho_s$ of cuprates places them in the renormalized classical regime, where, even for $J^\prime=0$, a N\'eel state would be present at $T=0$, as opposed to the quantum disordered regime, where fluctuations would kill the order parameter even at zero temperature.

The Heisenberg exchange interaction is obtained from the Hubbard model at half-filling in the $t$-$J$ approach.  Departing from half-filling, i.e. doping the parent compound with a holes density $h$, the measured magnetic transition $T_N(h)$ drops rapidly until a low temperature inhomogeneous magnetic order is established, often referred to as cluster spin glass (CSG), with onset temperature $T_f$. The CSG phase is the close precursor of the metal insulator transition, taking place at a critical density $h_s$ where superconductivity arises. It is a precursor both in the trivial sense that it lies in between the AF and the SC phases, and because increasing the temperature at any small finite density $h$ leads to metallic transport properties.\cite{AndoPRL2001} The qualitative understanding of the magnetic hole-doped ground state is that these holes localize on oxygen\cite{Zhang-RicePRB1988} at low temperatures and develop a large effective ferromagnetic (F) exchange $K>0$. The strong reduction of the transition, $dT_N/dh$, is then the effect of a magnetic frustration.\cite{AharonyPRL1988} A hole in the AF background actually produces a current around the plaquette, that can be mapped to a static spin twist\cite{GoodingPRL1991} and still described, in the hydrodynamic limit, by the \nsm, as it is done, e.g., in the spiral models.\cite{ShraimanPRL1989,LuescherPRL2007} Alternatively it may lead to diagonal or parallel stripes.\cite{MatsudaPRB2002}

The glassy nature of the CSG state \cite{JulienPB2003} is however controversial, since magnetic structure coexists with disorder\cite{MatsudaPRB2002,StockPRB2008}  and the observed features\cite{ChouPRL1995,WakimotoPRB2000} are not exclusive of a true spin glass state. We simply acknowledge that increasing hole density, hence frustration, leads to a more disordered state, whose experimental signature is the abrupt flattening of the magnetic transition dependence on hole density, i.e. $dT_f/dh \ll dT_N/dh$. In all cuprates this so-called CSG phase survives well inside the superconducting dome, in the form of low temperature {\em static} magnetism\cite{ChouPRL1993,BorsaPRB1995,BucciHypInt1997,NiedermayerPRL1998,SannaPRL2004,SaviciPRB2002} embedded in the metallic phase. The coexistence is intrinsic since it takes place over a range of hole densities $h-h_s\lesssim 0.02-0.03$, typically ten times greater than the inevitable hole inhomogeneity of stoichiometric origin.

A peculiar feature of the AF low doping region is the crossover between two thermal regimes, identified by the re-entrant behavior of the magnetic moment in \lsco, \ybcoy\ and \ycabco. $^{139}$La Nuclear Quadrupole Resonance (NQR)\cite{BorsaPRB1995} and muons\cite{BorsaPRB1995,BucciHypInt1997,NiedermayerPRL1998,SannaSSC2003,SannaPRL2004,ConeriPB2009} identify a thermally activated regime where the static moment is strongly reduced proportionally to hole density and a re-entrant regime where the moment recovers a nearly hole-independent zero temperature value. The crossover temperatures qualitatively agree with the onset of spin dynamics freezing, detected e.g. by Cu NQR\cite{JulienPRL1999} and muons,\cite{NiedermayerPRL1998,SannaSSC2003} but also with the activation temperatures of the variable range hopping regime in transport.\cite{KastnerRMP1998,AndoPRL2001,WangOngPNAS2001} Neutrons observe a decreasing static moment at $(\frac 1 2\, \frac 1 2\, l)$ in the re-entrant state,  associated with an increase of the rod $(\frac 1 2\, \frac 1 2\, q)$ {\em elastic} intensity\cite{EndohPRB1988} ($l$ integer, $q$ real) that matches the muon moment recovery. It is worth stressing that in these conditions it is improper to consider only the moment deduced from the Bragg peak intensity at $(\frac 1 2\,\frac 1 2)$, as if a simple N\'eel state were present. Partial and full sum rules\cite{LorenzanaPRB2005} are hard to implement and imply very large errorbars, whereas local probes offer a more direct, almost model independent way of determining the full static moment.

Besides these commonalities, remarkable differences are experimentally detected among the cuprates. We focus here on a few relevant points, limiting ourselves to the cuprate families stemming from the two parents \lco\ and \ybco\ (in this order, unless otherwise specified, for the values in parentheses): {\em i)} Distinct $T_N$;  {\em ii)} Distinct critical densities for the suppression of $T_N$ ($h_c\approx0.02,0.06$); {\em iii)} A region of pure CSG behavior, without superconductivity, sizeable\cite{AharonyPRL1988} in \lsco, almost vanishing in \ybcoy; {\em iv)} Distinct critical densities for the onset of the superconducting $T_c$ ($h_s\approx0.05,0.06$). The latter are rather close to each other, but clearly different,\cite{SushkovPRB2009} as it is confirmed also by Ca doping in \ybcoy, leading to $h_s=0.071$ for $8\%$ Ca, see Ref.~\onlinecite{SannaPRB2008}.
These main points  originate from the strong influence of the disordered heterovalent substituents (Sr$^{2+}$, Ca$^{2+}$), which place \lsco\ at the dirty limit and  \ybcoy\ at clean limit.\cite{DagottoNature2005,AlvarezPRB2005}


\section{\label{sec:results} Experimental results.}

\begin{figure}
\includegraphics[width=0.43\textwidth,angle=0]{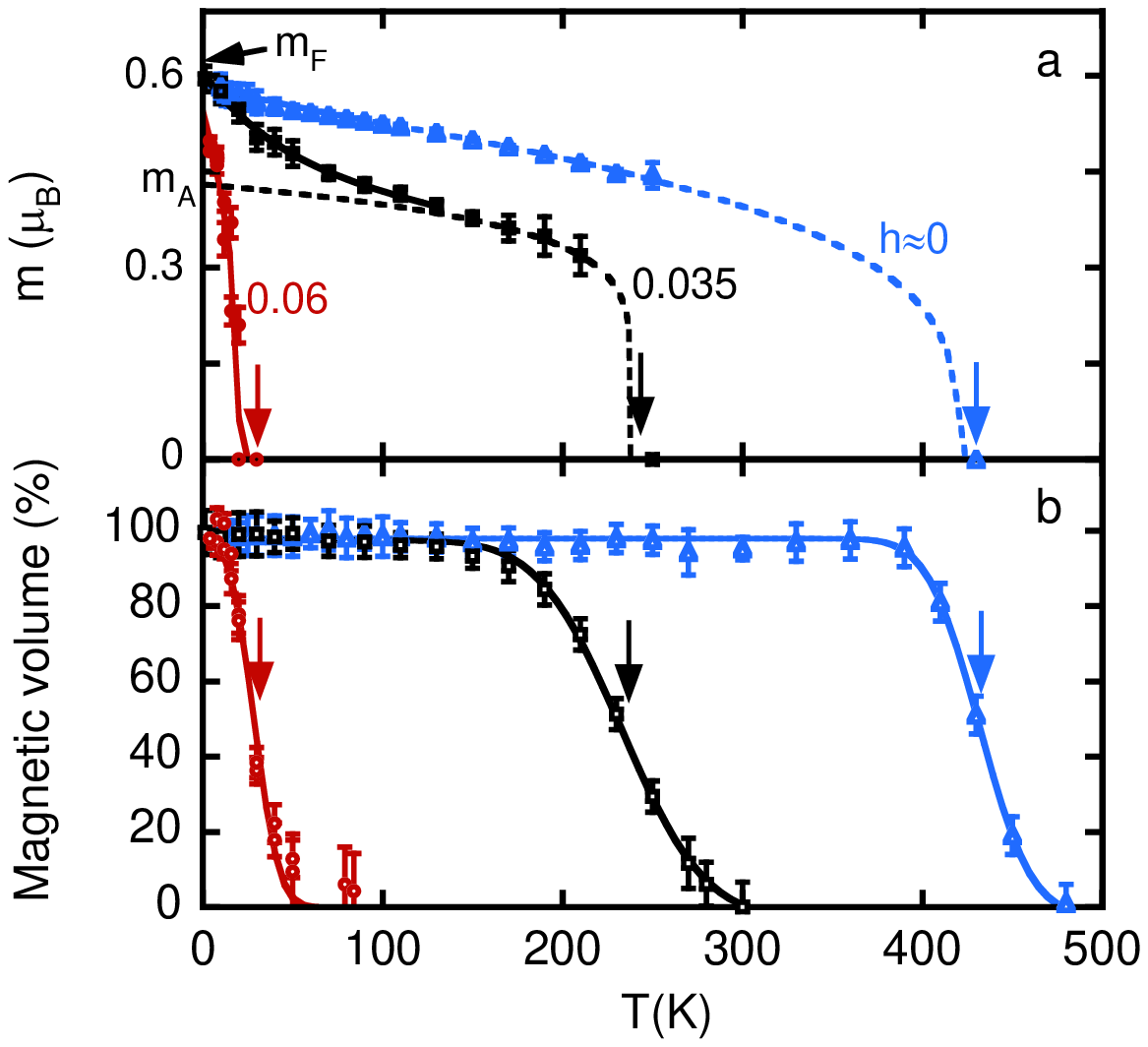}%
\caption{\label{fig:threemagnetizations} (color on-line) Temperature dependence for three \ybcoy\ samples of a) the magnetization (Eq.~\ref{eq:moment}) and b) the magnetic volume fraction defined in Sec.~\ref{sec:musr}. Vertical arrows indicate the magnetic transition.}
\end{figure}

Magnetic moments are proportional to the spin $S$ and mostly reside on Cu ions in the square plaquettes. In \ybcoy, as discussed in details in Appendix \ref{sec:localfield}, the internal field at the apical oxygen (AO) muon site provides a measure of the staggered moment $m(h,T)$ according to

\begin{equation}
\label{eq:moment}
m(h,T)=\mu_{2D} B_{AO}(h,T)/B_{AO}(0),
\end{equation}
where $\mu_{2D}\approx 0.6$ Bohr magnetons, is the value of the magnetic moment appropriate to the spin $1/2$ 2DHAF, with its quantum spin reduction\cite{SinghPRB1989} and $B_{AO}(0)$ is the muon internal field value at zero temperature in the undoped compound. This determination is independent of the details of the magnetic structure, provided that it preserves a local in-plane, two-sublattice collinear arrangement (Appendix \ref{sec:localfield}). Conservation of this local spin order condition allows muons to detect also the short range CSG state.

Figure \ref{fig:threemagnetizations}a shows the behavior of $m$  for three \ybcoy\ samples whose hole density ranges from nearly undoped, to just over the critical value where superconductivity appears, the strongly underdoped regime.
Figure \ref{fig:threemagnetizations}b displays the fraction of implanted muons detecting the magnetically ordered state, extracted from the muon asymmetry analysis (details in Appendix \ref{sec:musr}), which measures the fraction  of the sample volume belonging to the magnetic state.

First of all, the data of the two panels allow the determination of the magnetic transition temperature $T_m$ (we do not distinguish yet between $T_N$ and $T_f$, because the muon cannot {\em directly} identify the two states).
Secondly, at intermediate doping the order parameter $m$ deviates from a standard power law behavior like that of the undoped sample (dashed lines). The power law, with lower $T_m$ and rescaled magnetization, is followed only at high temperature, whereas an upturn (solid line) appears towards the {\em undoped} zero temperature value, $m(0)=1$.

\begin{figure}
\includegraphics[width=0.43\textwidth,angle=0]{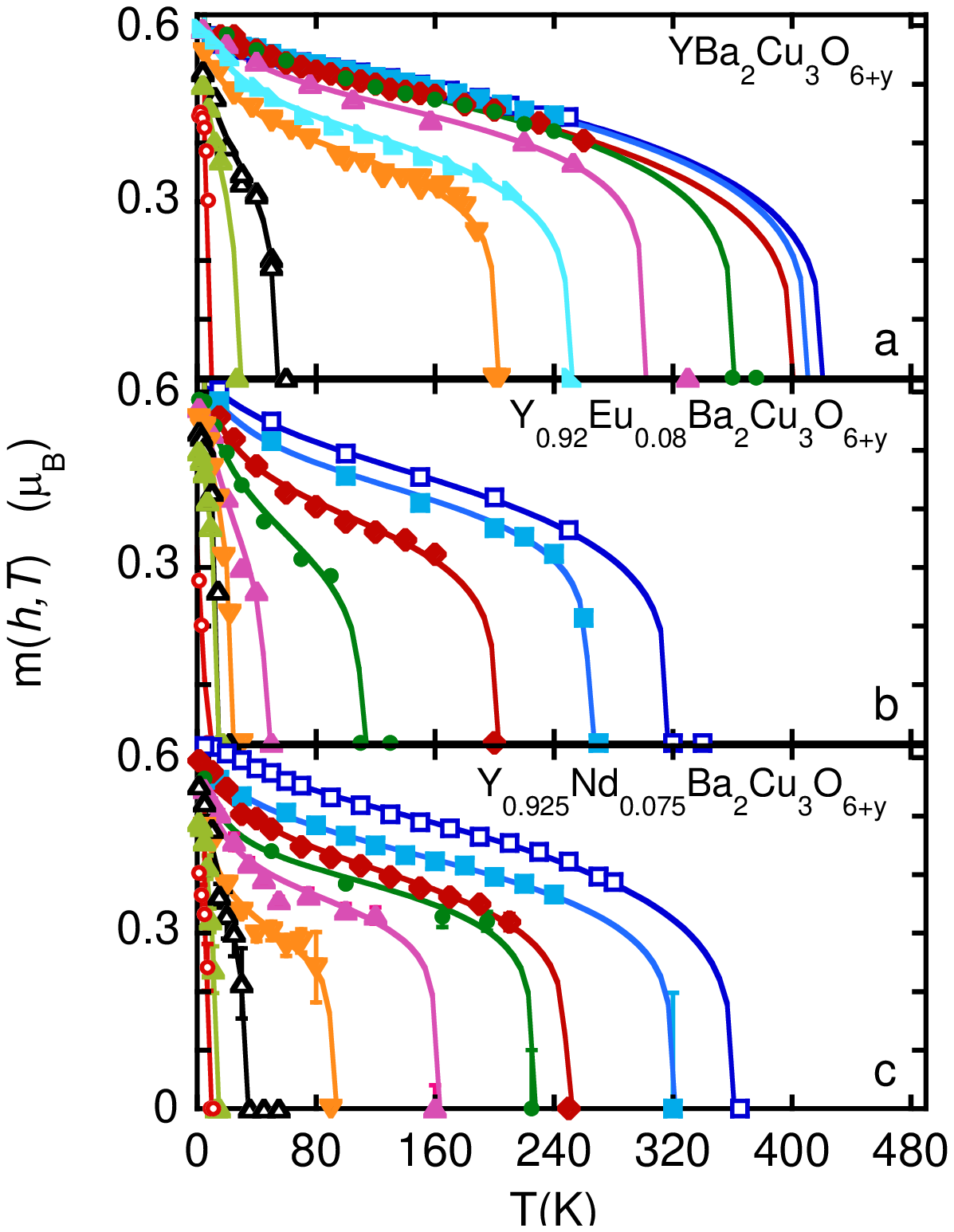}
\caption{\label{fig:fitTA} (color on-line) Temperature dependence of the moment $m$ with the best fit to Eq.~\ref{eq:fitTA}, for the three families.}
\end{figure}

For finite $h$ the recovery of the undoped magnetization $m(0)$ (Fig.~\ref{fig:threemagnetizations}a) justifies the term {\em re-entrance} for this behavior. A similar upturn is just barely perceptible also in the nearly undoped case of Fig.~\ref{fig:threemagnetizations}a;  it has been described before, first\cite{ChouPRL1993,BorsaPRB1995} in \lsco, then\cite{BucciHypInt1997,SannaSSC2003,ConeriPB2009} in \ybcoy\ and many more instances are shown in Fig.~\ref{fig:fitTA}. Muon and NQR agree quantitatively on the low temperature upturn, granting that this is {\em not} one of those very rare instances where the muon alters its local surroundings.

The data are naturally described as the crossover from  a low temperature magnetization towards a magnetically weaker high temperature regime, whose origin must be found in the thermal activation of charge and spin degrees of freedom. The functional form of the upturn looks like an exponential decay to the rescaled behavior. Thus, in order to distinguish the two regimes, we perform a preliminary fit of the data to the function\cite{SannaSSC2003}
$m(h,T)= [m_A + (m_F-m_A)e^{-c T}]$ $(1-T/{T_m})^\beta$,
where $m_{R,A}$ stand for the moment in the re-entrant and activated regimes, respectively. Notice that the former corresponds to the experimental ratio at $T=0$, $m_F=m(h,0)$ and the latter, $m_A$, yields the zero temperature extrapolation indicated by the dashed line in Fig.~\ref{fig:threemagnetizations}a. For $h>0.055$  we must revert to a simple power law fit
$m(h,T)= m_F \left(1-T/ {T_m} \right)^\beta$.
Similar fits are obtained for all samples, including those with Eu and Nd partial substitutions.

\begin{figure}
\includegraphics[width=0.5\textwidth,angle=0]{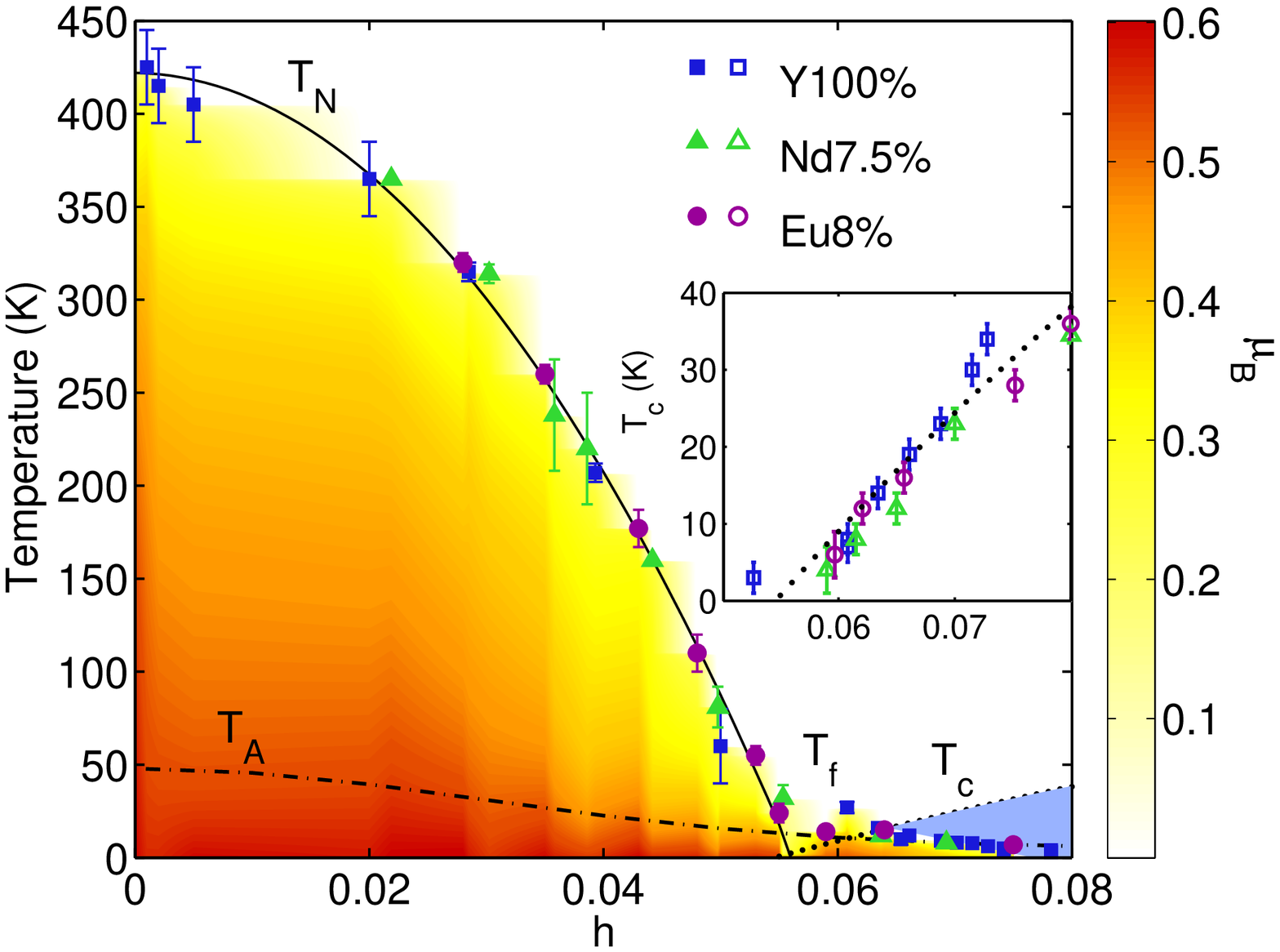}%
\caption{\label{fig:phasediagram} (color on-line) Common phase diagram of \ybcoy, \yeubco\ and \yndbco: transition temperatures $T_N$ and $T_f$ from the fits to Eq.~\ref{eq:fitTA}; experimental moment $m$ from fit and Eq.~\ref{eq:moment}, encoded in the nonlinear color mapping (see right bar); hole density $h$, see Sec.~\ref{sec:experimental} for details. The phase boundaries are best fits with $h_c=h_s=0.056(2)$ (see text) and $T_A$ labels the guide for the eye of Fig.~\ref{fig:TA}. Inset: zoom of the superconducting $T_c$, from magnetic susceptibility. }
\end{figure}

A compact view of the transition temperatures, with the moment as a color map is provided by Fig.~\ref{fig:phasediagram} for all these fits, for the three families of cuprates under investigation. It is evident that they display a unique behavior, within experimental errors (by the way, this observation is remarkably confirmed in all the data presented in this section).
Another straightforward consideration is that $T_m$ agrees qualitatively with many previous observations\cite{BrewerPRL1988,TranquadaPRB1989,Rossat-MignodPB1990,SannaPRL2004,MillerPRB2006} on unsubstituted \ybcoy, but only here it is presented as a function of calibrated hole density $h$. We recognize two regions: the first for $0<  h \lesssim 0.05$ where $T_m$ drops rapidly with increasing hole density, and we identify it with the N\'eel transition; the second for $h\gtrsim 0.05$, where the  $T_m(h)$ dependence is much weaker, and we associate this transition with $T_f$. The N\'eel temperature follows a quadratic behavior vs. hole density $T_{N0} (1-(h/h_{c})^2)$ with $T_{N0}=422(5)$ K and critical concentration $h_{c}=0.056(2)$ as shown by the solid line in Fig.~\ref{fig:phasediagram}.

\begin{figure}
\includegraphics[width=0.33\textwidth,angle=0]{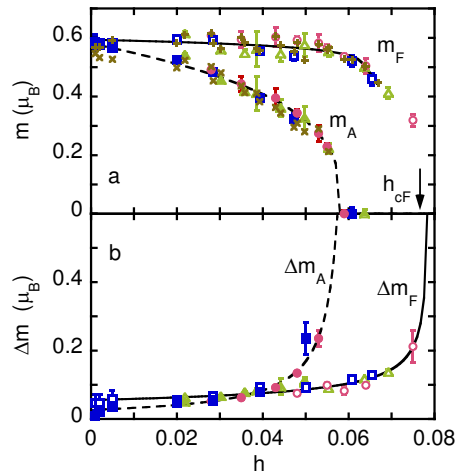}
\caption{\label{fig:mDm} (color on-line) Dependence on hole doping of: a) Re-entrant moment $m_F$, and activated moment $m_A$ from the fit to Eq.~\ref{eq:fitTA} (plus and cross, same quantities from the preliminary fit, see text); b) Width of the moment distribution measured by muons above and below $T_A$. Squares for Y100\%, circles for Eu8\% and triangles for Nd7.5\%.}
\end{figure}

The color map used in Fig.~\ref{fig:phasediagram} is nonlinear to emphasize the upturn of $m(T)$ (Fig.~\ref{fig:threemagnetizations}), identifying a crossover between the two regimes.  The re-entrant region is characterized by a uniform dark red color towards $T=0$, signifying that $m_F=m(h,0)$ approaches the full 2D moment $\mu_{2D}$ at all dopings, even underneath the superconducting dome. This is shown by the parameters $m_{F,A}$ plotted in Fig.~\ref{fig:mDm}a (crosses). The re-entrant zero temperature parameter $m_F$ is very close to one, up to the highest hole densities, until it undergoes an abrupt reduction around a distinct $h_{cF}\approx0.08$.  Conversely the activated parameter $m_A$, plotted in the same figure, displays a clear distinct critical behavior, fitted to a power law $m_A=(1-h/h_{c})^\alpha$ with $\alpha=0.35(5)$ and $h_{c}=0.058(2)$, equal within errorbars to the value of Fig.~\ref{fig:phasediagram}.

It is tempting to include the $m_A(h)$ power-law behavior of Fig.~\ref{fig:mDm} directly into the fit model, identifying its doping dependence as $m_A(h) = m_F(h)\left(1- h/ {h_c} \right)^\alpha $. It is however clear that a constant hole density does not reproduce the thermally activated features of $m$ and a straightforward extension is to replace $h$ by  $h_A(T)=h \exp(-T_A/T)$, assuming for the sake of the argument that delocalized (hopping) holes are responsible of the magnetization reduction. This leads to

\begin{eqnarray}
\label{eq:fitTA}
m(h,T) &=& m_A(h,T)\left(1-\frac T {T_m} \right)^\beta\cr
&=& m_F \left(1 - \frac h {h_{c}} e^{-T_A/T}\right)^\alpha \left(1-\frac T {T_m} \right)^\beta
\end{eqnarray}

This activated-hole function yields excellent best fits of the data as it is shown by the solid lines in Fig.~\ref{fig:fitTA}.  We note however that Eq.~\ref{eq:fitTA} could  also hold in a stripe scenario\cite{ZaanenPRB1989}, where it is possible to imagine a mechanism by which the influence of thermally activated excitations on magnetism depends on the stripe density, which in turns is proportional\cite{PoilblancPRB1989} to $h$. The difference between Eq.~\ref{eq:fitTA} and the  functional form of the preliminary fit is often beyond resolution. The critical temperatures ($T_N$ and $T_f$) fitted in the two models are almost identical, so that Fig.~\ref{fig:phasediagram} is unchanged. The new parameter $T_A$, is shown in Fig.~\ref{fig:TA} and its guide for the eye is reported also in Fig.~\ref{fig:phasediagram} as a quantification of the crossover temperature (dashed line). The activation energy follows a dome that replicates roughly  the AF phase boundary, albeit with a large down-shift. On entering the CSG phase coexisting with superconductivity, for $h>h_c$, the fit can still be performed, but $T_A$ is hardly distinguishable\footnote{\protect{A single sample, Y100\% with $h=0.06\gtrsim h_c$, stands just outside the limits of this classification, since its magnetic transition, $T_m=28$ K is larger that its activation temperature $T_A=15$ K .}} from $T_f$, hence we report only the latter in Fig.~\ref{fig:phasediagram}.

The two moments displayed in Fig.~\ref{fig:mDm}a have a counterpart in the activated-hole model (Eq.~\ref{eq:fitTA}): the zero temperature moment, $m_F$, is still a free parameter, whose functional dependence on $h$ is directly determined by best fit procedures, whereas $m_A$ is obtained as $\lim_{T\to\infty} m_A(h,T)$. These new determinations (colored symbols) agree perfectly with the previous ones (crosses) as it is shown in the figure.

Two very distinct critical behaviors are also shown in panel b, where we plot the widths $\Delta m_{F,A}$ of the two static moment distributions. These widths are obtained directly from the static relaxation of the muon asymmetry precession as $\sigma_{AO}(T)/2\pi\gamma B_{AO}(T)$ (Sec.~\ref{sec:musr}), taking cuts through the data at two temperatures, the lowest one, 2 K, below all $T_A$ and another one, 70 K, above the largest $T_A$. Both quantities display a divergence which points, within errors, to the same two distinct critical densities  mentioned above, $h_{c}$ and $h_{cF}$.

\section{\label{sec:discussion}Discussion}
\subsection{\label{subsec:distortions} Isovalent Y substitutions}

Disorder must be considered as an additional parameter controlling the phase diagram of oxides, distinguishing a clean and a dirty (heavily disordered) limit.\cite{DagottoNature2005,AlvarezPRB2005} The latter case is clearly represented\cite{NiedermayerPRL1998,ConeriPB2009,SannaArXiv2009} by \ycabcoy\  and by \lsco. These examples consist of heterovalent cationic substitution in layers nearest neighbor to CuO$_2$, determining a random distribution of Coulomb impurity potentials accompanied also by local structural distortions. The strong influence of the latter on the electronic properties\cite{DabrowskiPRL1996} is specifically demonstrated by the emergence of static stripes, evoked in \lsco\  by the La-Nd,Eu\cite{TranquadaNature1995,KlaussPRL2000} partial substitution around the hole concentration $h=1/8$ and by the shift of the MI transition in heavily Y-substituted\cite{LutgemeierPC1996} \ybcoy.

\begin{figure}
\includegraphics[width=0.43\textwidth,angle=0]{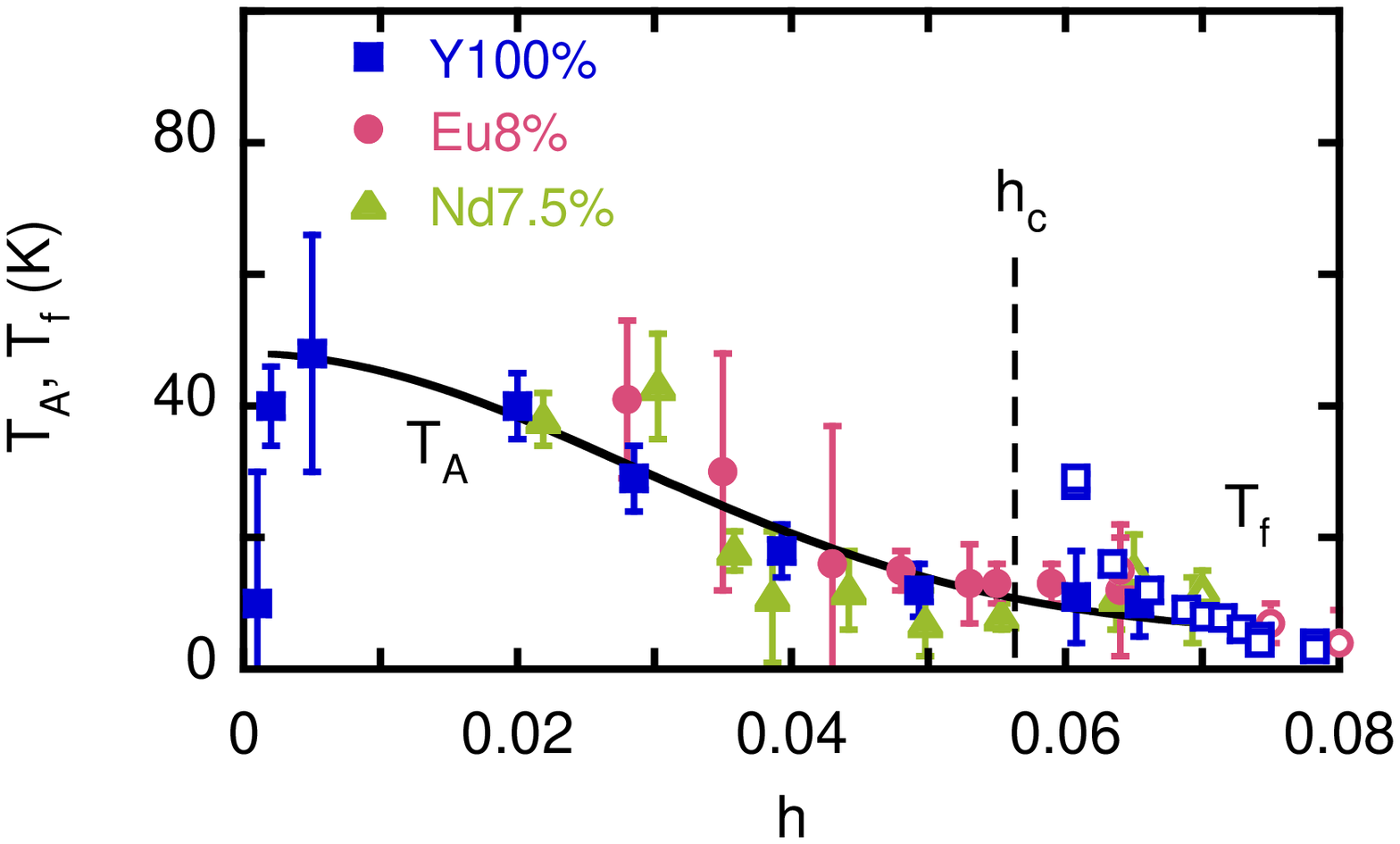}
\caption{\label{fig:TA} (color on-line) Dependence on hole doping of the activation temperature $T_A$ from the best fits to Eq.~\ref{eq:fitTA}. The line is a guide for the eye and the data for $h>h_C$ represent the transition temperature $T_f$.}
\end{figure}

In order to disentangle structural and Coulomb disorder we compare isovalent (Eu,Nd) and heterovalent (Ca) substitutions. Based on the ionic radii\cite{Shannon} - Y$^{3+}$ (101.9 pm), Eu$^{3+}$ (106.6 pm), Nd$^{3+}$ (110.9 pm), Ca$^{2+}$ (112 pm) - we evaluate the cation mismatch via the radius standard deviation, $\sigma_r=1.3, 2.4, 2.2$ pm for  Eu8\%, Nd7.5\%  and \ycabcofive\ (Ca5\%), respectively.
Our data show that the N\'eel  and superconducting transitions  of the Nd and Eu substituted samples are identical within experimental error to those of the pure compound (Fig.~\ref{fig:phasediagram}), whereas Ca5\%, with the same $\sigma_r$ as Nd7.5\%, displays a markedly different behavior.\cite{SannaPRB2008,SannaArXiv2009}
We conclude that, for dilute Y replacements, isovalent disorder is by far less effective than heterovalent disorder.

As a consequence the three families (Y100\%, Eu8\% and Nd7.5\%) equivalently identify the cuprate system closest to the clean limit and in the discussion of the following subsections we shall refer generically to them as the YBCO system.

\subsection{\label{subsec:ltphase} Nature of the two magnetically ordered regimes}
Our data establish a coherent and systematic picture of the AF YBCO phase which is characterized by two distinct regimes, re-entrant and thermally activated, in agreement with previous indications\cite{BorsaPRB1995} on the magnetic behavior of \lsco. The low temperature re-entrant regime witnesses the recovery of the full 2DHAF moment, $\mu_{2D}$, as it is shown in Fig~\ref{fig:mDm}a (solid curve). This finding is reconciled with the opposite sign anomaly in the neutron magnetic Bragg peaks\cite{Rossat-MignodPB1990,TranquadaPRB1989} considering that the neutron elastic intensity increases\cite{TranquadaPRB1989} at low temperatures along the 2D ridges (for $\bm q \parallel \hat {c^*}$). Although the total elastic scattering does not allow a unique identification of the low temperature magnetic structure, the incommensurate peaks\cite{MatsudaPRB2002} observed in \lsco\ above $h_c=0.02$ are tentatively interpreted in terms of a spiral spin state.\cite{LuescherPRL2007} In \ybcoy\ the experimental evidence of incommensurate order is missing, but similar magnetic correlations are detected.\cite{StockPRB2008} Stripes or spiral spin structures are indeed a viable candidates since they are locally collinear, in-plane and staggered, as required by the muon local field (Appendix \ref{sec:localfield}) and they can provide the large local moments that we detect.

Another important feature of the re-entrant state is the narrow width of the observed moment distribution $\Delta m_F$, that reaches the value 0.4 only close to the critical point $h_{cF}$, as shown in Fig.~\ref{fig:mDm}b. Notice that narrow widths and staggered in-plane collinearity  extend also to the CSG phase coexisting with superconductivity, for $h_{c}\le h \le h_{cF}$. This is not what is expected of a canonical spin glass, where muons typically\cite{UemuraPRB1985} measure $\Delta m\approx 1$ and randomly oriented local fields.

\begin{figure}
\includegraphics[width=0.43\textwidth,angle=0]{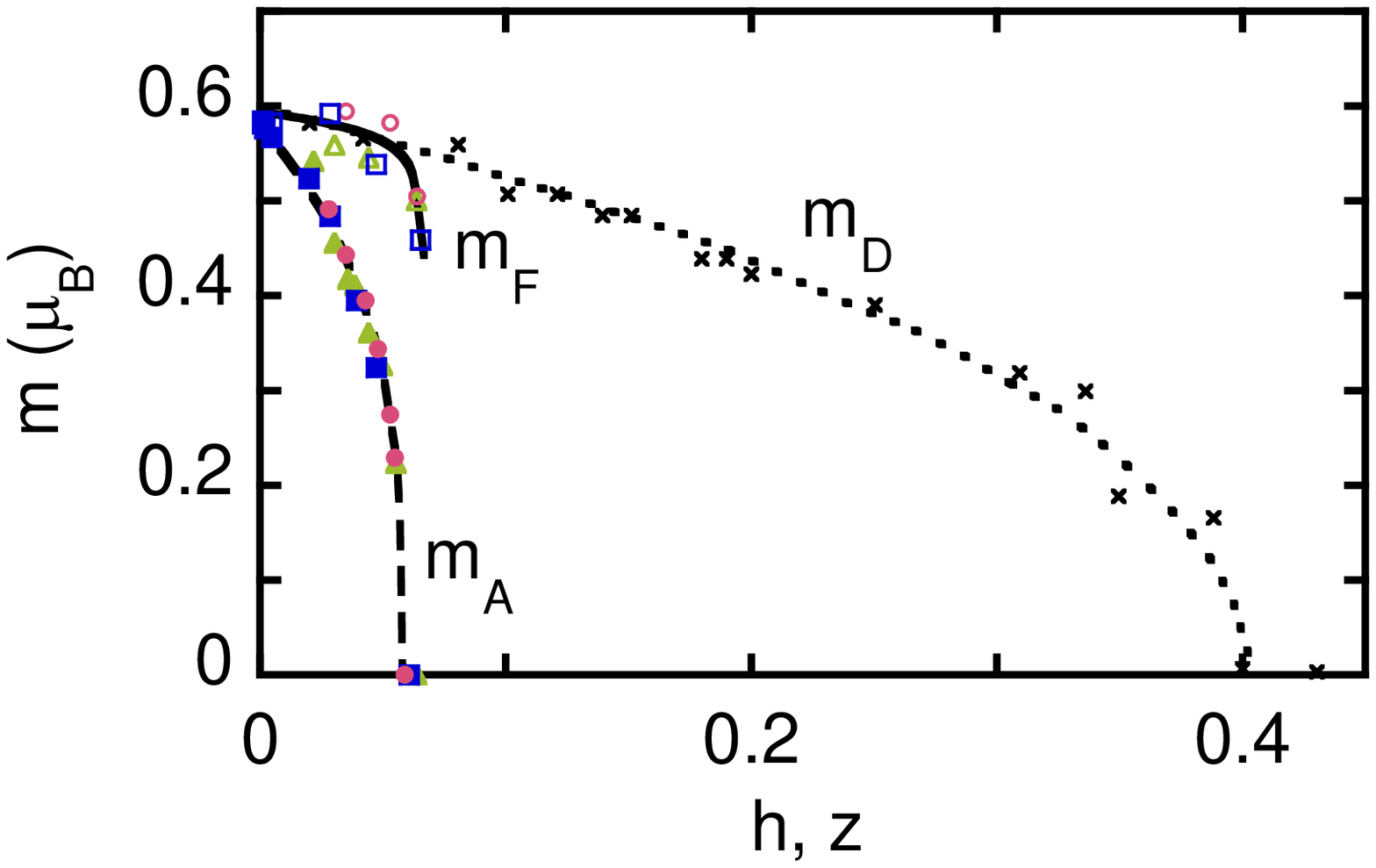} 

\caption{\label{fig:dilution} (color on-line) Magnetic moments $m_F$ and $m_A$ (same as Fig.~\ref{fig:mDm}) compared with theory\cite{LiuPRL2009} (dotted line) and experiments\cite{CarrettaPRB1997} on \lscozn\ (crosses).}
\end{figure}

The main signature of the CSG phase is the change from a rather large slope of the transition temperature, $dT_N/dh$, below $h_{c}$, to a very modest slope $dT_f/dh$ above. Notice that the $T_A$, the activation  temperature defined by Eq.~\ref{eq:fitTA}, shares this same modest slope, and the two temperatures, $T_A$ and $T_f$ follow the same curve in Fig.~\ref{fig:TA}. This may be simply understood in terms of the disappearance of the high temperature, activated magnetic regime, above the critical density $h_{c}$, while the magnetic ground state detected in the re-entrant regime survives as shown by the continuous behavior of $m_F(h)$ and $\Delta m_F(h)$ across $h_c$, Fig.~\ref{fig:mDm}. This leads us to identify the same ground state in the CSG phase coexisting with superconductivity ($h_c\le h\le h_{cF}$) and in the re-entrant regime of the antiferromagnet, a common state that we propose to dub frozen antiferromagnet (FAF), in contrast with the thermally activated antiferromagnet, henceforth TAAF, established  only above the crossover temperature, for $h<h_c$.

Insight on the nature of the FAF state
is provided by Fig.~\ref{fig:dilution} which shows that $m_F$ is actually following the initial slope typical of magnetic site dilution in cuprates\cite{LiuPRL2009} (dotted line). A similar conclusion can be derived from recent low temperature muon results\cite{OferPRB2008} in \calabalacuo. One way to look at this is to suppose that in the re-entrant regime holes are localized and each of them in first approximation cancels one Cu spin, leading to magnetic dilution with $z=h$.

The dilution theory has recently found\cite{LiuPRL2009} excellent agreement with available \lscozn\ data\cite{CarrettaPRB1997,VajkS2002,VajkSSC2003} (crosses in Fig.~\ref{fig:dilution}), by taking into consideration a small additional frustrating effect of the local Zn impurity that slightly reduces the effective coupling. Also localized holes give rise to a frustration,\cite{HasselmannPRB2004} in this case of dipolar nature, that is often indicated as the cause of the dramatic reduction of $T_N(h)$. Our $m_F(h)$ data in Fig.~\ref{fig:dilution}, however, follow very closely the dotted curve of Ref.~\onlinecite{LiuPRL2009} until the magnetic component disappears, showing that the leading phenomenon at low temperature is moment dilution and not frustration.  This result indicates that holes, either by incoherent freezing or by self-organization produce only a rather moderate frustration in the AF background of the FAF state.

\begin{figure}
\includegraphics[width=0.43\textwidth,angle=0]{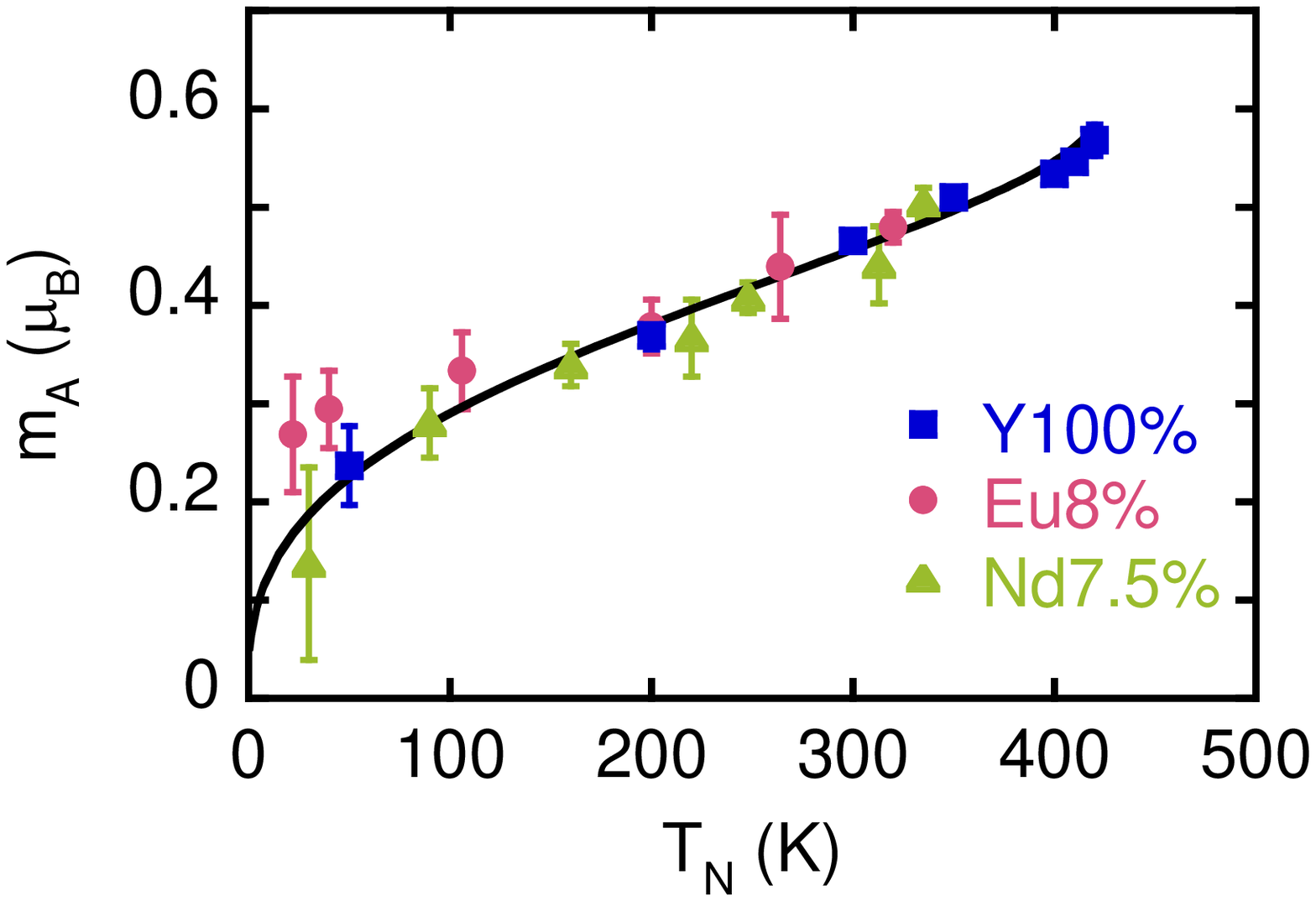}

\caption{\label{fig:mATN} (color on-line) Scaling of the moment in the activated regime, $m_A$, with $T_N$ (for the solid line see the text).}
\end{figure}

Conversely the slope of the thermally activated moment $m_A(h)$ is much steeper than the predictions of the dilution theory, even at extremely low doping, as the dashed and dotted curves show in Fig.~\ref{fig:dilution}. It is noteworthy that the vanishing of $m_A$, the divergence of $\Delta m_A$ (Fig.~\ref{fig:mDm}) and the suppression of $T_N$ (Fig.~\ref{fig:phasediagram}), all occur at the same critical density $h_c$, thus linking all these properties to the specific TAAF phase alone.

Figure \ref{fig:mATN} shows that the activated moment $m_A(h)$ and $T_N(h)$ scale roughly linearly with each other over a large range of $h$ (the solid line is obtained by plotting the previously defined functions $m_A(h,\infty)$ vs. $T_N(h)$ with $h$ as an implicit parameter). It implies that both $m_A$ and $T_N$ scale with an effective exchange coupling $J_{eff}(h)$, strongly reduced by the activated process, either by finite-size effects\cite{BorsaPRB1995} or by a very large frustration,\cite{AharonyPRL1988,CherepanovEPJB1999} much larger than that present in the re-entrant regime.

The magnetic moment in the TAAF state follows the phenomenological fit of Eq.~\ref{eq:fitTA}. Nominally, this fit  justifies the low temperature regime in terms of hole localization. Here the muon is directly sensitive only to spin dynamics, hence, recalling that spin/charge separation may take place in cuprates,\cite{AndersonS1987} we cannot actually prove {\em directly} that charges freeze as well. Freezing processes with similar low energy scales are detected also by \msr\cite{NiedermayerPRL1998,SannaSSC2003} and NQR\cite{SuhPRL1998,JulienPRL1999} $T_1^{-1}$ relaxations (the latter being sensitive also to charge dynamics) in \lsco\ and in \ybcoy\ as well.

Direct evidence that thermal activation regards also charge degrees of freedom comes from transport, where low temperature variable range hopping\cite{AndoPRL2001,SunPRL2004} is detected both in dirty \lsco\ and in clean \ybcoy, even at very low hole densities.\cite{WangOngPNAS2001} The energy scale of charge activation is comparable to the one we observe here (resistivity at the lowest doping in Ref.~\onlinecite{WangOngPNAS2001}, $T_N=380$ K, shows a low temperature divergence starting below 70 K). The TAAF phase is therefore the same that strongly violates the Mott-Ioffe-Regel limit (Fermi wave vector times mean free path much less than one), but still displays a metallic transport character\cite{AndoPRL2001} even at very low hole densities. The coincidence of the onset of this so-called {\em bad metal}\cite{EmeryPRL1995} state, with the reduction of the magnetic moment around $T_A$ consolidates the association of thermal activation with charge carriers implied by Eq.~\ref{eq:fitTA}. This point deserves a systematic investigation by transport in very low doped single \ybcoy\ crystals, scarcely studied to date.

Notice that when the TAAF phase disappears, for $h> h_c$, the samples become electronically inhomogeneous, since the well known nanoscopic phase separation\cite{SannaPRL2004} takes place. In this case the {\em bad metal} state exists already at $T=0$ in the percolating superconducting background, but inside the interspersed nanoscopic magnetic clusters it develops only above $T_f$.

If thermally activated holes determine directly the reduction of $m$ in the TAAF state they must be responsible of a much larger frustration than independently localized holes. Such a view could be reconciled with recent extensions of the spiral model\cite{LuescherPRL2007}, where increasing hole localization lengths, $\kappa^{-1}$, are shown to reduce  the staggered moment very effectively. The same approach successfully accounts for activated hole transport\cite{JuricicPRL2004,JuricicPRB2005}, i.e.~for the effect of spin texture on carrier mobility, but none has yet taken in due consideration the reverse effect of hole motion on spin texture, which is very relevant, as we show.
The two-regime-behavior of Fig.~\ref{fig:fitTA} would then suggest that the localization length is not temperature independent in cuprates:  while a rather large length $\kappa^{-1}$ (cfr. the value of 100 pm quoted in Ref~\onlinecite{LuescherPRL2007} for \lsco) is appropriate for the thermally activated regime, a smaller length would be characteristic of the FAF state. Such a temperature-dependent localization scale suggests that at very low temperature the effects of charge localization in a lightly doped Mott-Hubbard insulator are not fully captured by the standard impurity model of doped semiconductors investigated so far in the literature.

However, as we have noted below Eq.~\ref{eq:fitTA}, the fit of the muon measurements to this function by itself does not rule out alternative stripe based ideas.  For instance a similar connection with  thermally activated hole density could emerge considering stripe domains of finite length and with rather large $d=1/2h$ separation\cite{LorenzanaPRL2002} at the low densities considered here. In these conditions metallic 1D stripes would not percolate across the domains, whereas the antiferromagnetic regions would. Thermally activated hopping would then switch the system to the opposite condition of percolation for transport and separation of independent magnetic clusters,  at the same time. This process is already favored by the reduced effective exchange across the metallic 1D stripe, and it  would lead to finite-size reduction of both $m$ and $T_N$.

Summarizing, both spirals and  stripes could qualitatively justify the YBCO \msr\ results, but only further theoretical work can determine if either of these hypotheses reproduces the observed crossover and the associated large change in the static local moment.

Another important result of the present work is that the critical value, $h_c=0.056(2)$, found for the suppression of $T_N(h)$ and $m_A(h)$  coincides with the critical value for the onset of $T_c(h)$, $h_s=0.056(2)$. Both values agree within our errorbars with a recent very careful single crystal determinations\cite{BrounPRL2007} of $h_s=0.0539(1)$. According to our data also $T_N(h)$, in the YBCO clean limit, follows a parabolic behavior analogous to that\cite{TorrancePRB1989,LiangPRB2006} of $T_c(h)$ and very different from theoretical predictions.\cite{CherepanovEPJB1999} Notice that older systematic results,\cite{BrewerPRL1988,Rossat-MignodPB1990} reported only vs.~\-oxygen content, without hole doping calibration, could not provide such information.
Therefore our results indicate that if superconductivity and the TAAF state were the only competing orders, $h_c=h_s$  would represent a true quantum critical point (QCP). The critical density is rather lower than that more often indicated for a QCP in cuprates\cite{SachdevRMP2003}, but other authors\cite{PanagopoulosPRB2005} have considered a similar additional location.  The real-world system however does not develop the QCP, because it switches instead to the FAF  ground state throughout a wide doping range,
from $h\approx 0$ to well past the onset of superconductivity. Additional evidence on this point comes from our parallel work on the heterovalent substitutions.\cite{SannaArXiv2009}

\section{\label{sec:conclusion} Conclusions}

We have extensively investigated YBCO in the  antiferromagnetic low doping region as a model for the clean limit cuprate. Since the influence of disorder induced by Nd and Eu is found to be negligible, we extract the parameters of the low temperature re-entrant and of the thermally activated regimes from the whole set of data. The crossover between these two regimes is undoubtedly associated with the closely related thermal activation of both spin and charge degrees of freedom. This is implemented in a single phenomenological model for the temperature and doping dependence of the moment $m(h,T)$, which fits our data throughout the entire investigated range.

Our analysis leads to the identification of a common ground state for very different doping regimes at the two sides of the metal-insulator transition: the so-called Cluster Spin Glass phase and the re-entrant antiferromagnetic phase. We dub this unique state a frozen antiferromagnet.

We have shown that in the frozen antiferromagnetic state the ordered magnetic moment is subject to negligible frustration and it follows the same trend of magnetic site dilution. Conversely the fast drop of $T_N(h)$ and of the staggered moment, $m_A(h)$ are characteristic of the thermally activated antiferromagnetic phase.

Finally, our data indicate that the main boundaries in the clean-limit phase diagram, describing the two main order parameters, magnetic and superconducting, both follow parabolic curves\cite{TallonPRB1995,LiangPRB2006,SannaPRL2004,SannaPRB2008} sharing a zero temperature intersection at $h=h_c=h_s$. At face value this could mean that $h_c=0.056$ is a quantum critical point for the  cuprate clean limit. The QCP is then superseded by the FAF behavior of the real-world compound at low temperatures.

\appendix
\section{\label{sec:localfield}The muon local field}

\begin{figure}
\includegraphics[width=0.43\textwidth,angle=-90]{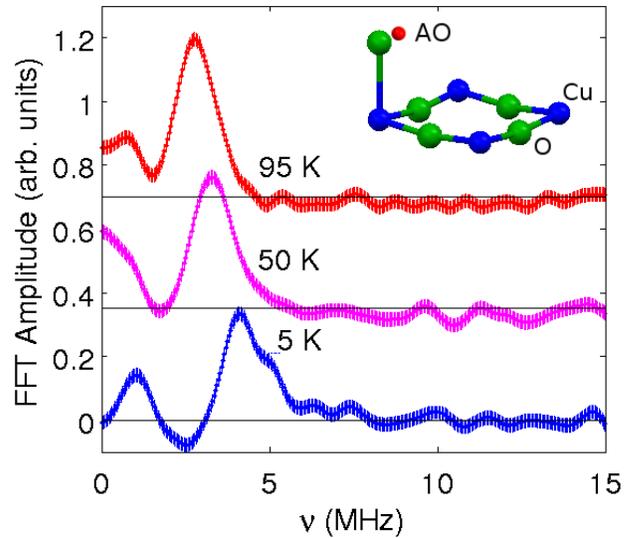}%
\caption{\label{fig:FFT} (color on-line) \ybcoy, $y=0.27 ,\, h=0.023$: FFT amplitude of the precessing asymmetry a three temperatures, vertically shifted for clarity. Inset: muon AO site, with apical oxygen and the CuO$_2$ plaquette.}
\end{figure}

Three types of muon stopping sites are distinguished in cuprates. The dominant one, observed in all compounds, irrespective of both the details of the perovksite structure and doping, is bound to the apical oxygen (AO), just above or below the CuO$_2$ plaquette (inset of Fig.\ref{fig:FFT}). The low temperature value of the field probed by muons at this site is $B_{AO}^L=40$ mT for \lco, Ref.~\onlinecite{WeidingerPRL1989}, and $B_{AO}^Y=30$ mT for \ybco, Ref. \onlinecite{BrewerPRL1988}, respectively.  A second site is specific of \ybcoy\ for $y> 0.2$ and it is attributed to muons bound to chain oxygen (CO), with roughly half the field intensity. A third site, directly bound to the oxygen ions of the plaquette (PO), with large internal fields, is observed only\cite{Brewer:private} in undoped \ybco.

The AO site is the main subject of our analysis, and its assignment is based on numerous observations.\cite{GlucklerPC1989,NishidaHypInt1990,WeberHypInt1990,PinkpankPhysC1999} One of them is that the value of $B_{AO}^Y$ at zero temperature and doping is reproduced by simple dipolar sums: if $S(h,T)$ is the average staggered Cu spin defined in Sec. \ref{sec:results}, $\hat S_i$ is the spin direction at site $i$, $\bm{r}_i=r_i\hat r_i$ is the vector joining the muon site to the $i$-th Cu spin and $g$ is the Land\'e factor, we can write

\begin{eqnarray}
\label{eq:muonfield}
B_{AO}(h,T)&=&\frac {\mu_0  g \mu_B}{4 \pi} \left|\sum_j \frac {3(\hat S_j\cdot \hat r_j)\hat r_j + \hat S_j }{r_j^3}\right| S(h,T)\cr
&=&\left| \bm{D} \right| S(h,T),
\end{eqnarray}

The zero temperature value is reproduced in this site with the standard\cite{SinghPRB1989} 2D spin reduction, $gS(0,0)=0.6$. A small isotropic super-hyperfine contribution cannot be a-priori excluded, but it would preserve the validity of Eq.~\ref{eq:muonfield}, with the simple addition of a term $A\hat S_1$ to $\bm D$, $\bm S_1$ being the Cu spin nearest to the muon.

The sums defining vector $\bm{D}$ in Eq.~\ref{eq:muonfield} converge rapidly, thanks to the alternating sign of $\hat S_i$ in the AF structure (the staggered moment). Even in the very underdoped superconductor YBa$_2$Cu$_3$O$_{6.35}$, where in-plane magnetic correlation lengths are shortest,\cite{StockPRB2008} they never fall below $\xi_{a,b}\approx 12$ \AA, and the relative contributions to Eq.~\ref{eq:muonfield} for $r_j>\xi_{a,b}/2$ are only a few percent. Hence the local field may be written as a constant, $D=|\bm{D}|$, times $S(h,T)$

The rapid convergence of the dipolar sums and the tensorial nature of the  interaction guarantee that $D$ does not change as long as the local spin directions $\hat S_i$, nearest neighbors to the muon,  remain {\em  i)} collinear, {\em  ii)} parallel to the $ab$ plane, and {\em  ii)} staggered. These points follow from the consideration that the value of $B_{AO}$ obtained in the simple N\'eel arrangement is actually invariant under in-plane $\hat S_i$ rotations for a muon site coinciding with that of the apical oxygen. The AO site is displaced only by 110 pm from this symmetry position and the invariance is roughly preserved.
Conversely if either the moment were drastically reduced, or the local spin arrangement did change significantly from staggered, collinear and in-plane, the low temperature value of the local muon field of Eq.~\ref{eq:muonfield} would also change. For these reasons we conclude that the nearly constant local field $B_{AO}(h,0)$ (rescaled to $m_F(h)$ in Fig.~\ref{fig:mDm}a) requires both nearly constant $D$ and $S$. An unvarying $D$ in turns implies that a collinear staggered in-plane spin arrangement is at least locally preserved at low temperature for all hole densities.

The argument discussed above could in principle be questioned in the presence of an incommensurate magnetic structure with long, $h$-dependent wavelength, giving rise to broad distributions of local fields with a sharp peak at the Van Hove singularity. Since only this peak would be observed and fitted, any doping-dependent change could modulate not only the peak position, but also its spectral weight (i.e.~reduce the fitted initial muon asymmetry). However we always detect an unchanging spectral position and the full initial muon asymmetry, ruling out this possibility.

Let us conclude by reviewing the experimental evidence: the negligible variation of the low temperature AO field throughout the  magnetic phase diagram is amply discussed in Sec.~\ref{sec:results}. The field distribution itself does not change drastically as it is witnessed by the very similar low temperature asymmetries in the six panels of Fig.~\ref{fig:asymmetry}, spanning from $h=0.02$ (top) to $h=0.07$ (bottom). If local correlations did vary significantly we would observe the appearance of distinct frequency components, signalling different internal fields, whereas the only notable change is that the damping of the precessions (i.e. the width of the field distribution) increases with hole content, as shown also in Fig.~\ref{fig:mDm}b.

The field distribution does not change appreciably with temperature either. Figure \ref{fig:FFT} shows representative Fast Fourier Transform (FFT) spectra of the muon precessing asymmetry below, around and above $T_A$. They display two peaks corresponding to the AO and CO sites, which shift to lower frequency as $S(T)$ is reduced, preserving their shape.

\section{\label{sec:experimental}Experimental details}

\begin{figure}
\includegraphics[width=0.4\textwidth,angle=0]{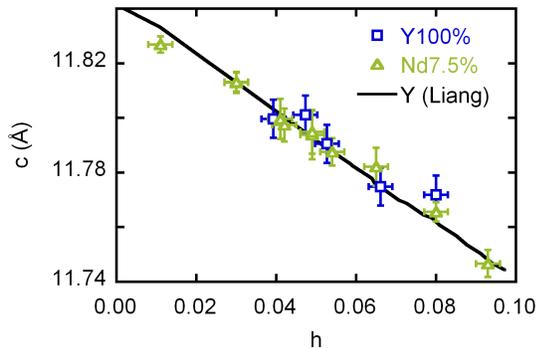}%
\caption{\label{fig:ch} (color on-line) Lattice parameter $c$ from XRD versus hole density from thermopower. The solid line is the calibration from Ref.~\onlinecite{LiangPRB2006}.}
\end{figure}

High quality polycrystalline samples of \ybcoy, \yndbco\ and \yeubco\ with $0.05(2) < y < 0.48(1)$ are obtained by sintering. The Y100\% series is the same of Ref.~\onlinecite{SannaPRL2004}. The absence of spurious phases is confirmed by systematic x-ray Rietveld refinements, and the oxygen content has been fixed by a topotactic technique, which consists in the oxygen equilibration of stoichiometric quantities of two end members ($y=0.05(2)$ and
$y=0.98(1)$), packed in sealed vessels with minimum free volume.\cite{MancaPRB2000,MancaPRB2001} Low temperature annealing produces homogeneous samples, with well ordered CuO chains in the basal plane, with an absolute error of $\delta y = \pm 0.01$ in oxygen content per formula unit (cross checked by iodometric titration and thermogravimetry on each sample). The mobile carrier content is obtained from
systematic measurements of the room temperature Seebeck coefficient,\cite{ObertelliPRB1992} with a calibration procedure \cite{SannaPRB2008} of the exponential
dependence of $S$ vs $h$, performed on empty-chain \ycabco\ samples. Figure \ref{fig:ch} shows that this calibration agrees very well with that extracted\cite{LiangPRB2006} from the lattice parameter $c$. Notice that our determination of a common hole doping density scale for all compounds is not  based on the a-priori assumption of a common equal doping density for optimal superconductivity $h_o\approx0.16$, as it was done for instance also in our previous work.\cite{SannaPRL2004,SannaJSNM2005}. Slight deviations between Fig.~\ref{fig:phasediagram} and these previous published Y100\% data are due to the different calibration.

Superconducting behavior characterizes samples with $h>h_s=0.056$, whose critical temperature $T_c$ is determined from the linear extrapolation of
the 90\% to 10\% drop in the diamagnetic susceptibility measured in a field of $\mu_0 H = 0.2 mT$.

 The $\mu$SR experiments were performed on the MUSR and EMU spectrometers at the ISIS Facility of the Rutherford Appleton Laboratory (UK) in longitudinal geometry with zero external field.\cite{Schenck1986}

\section{\label{sec:musr}Zero-Field $\mu$SR analysis}

\begin{figure}
\includegraphics[width=0.43\textwidth,angle=0]{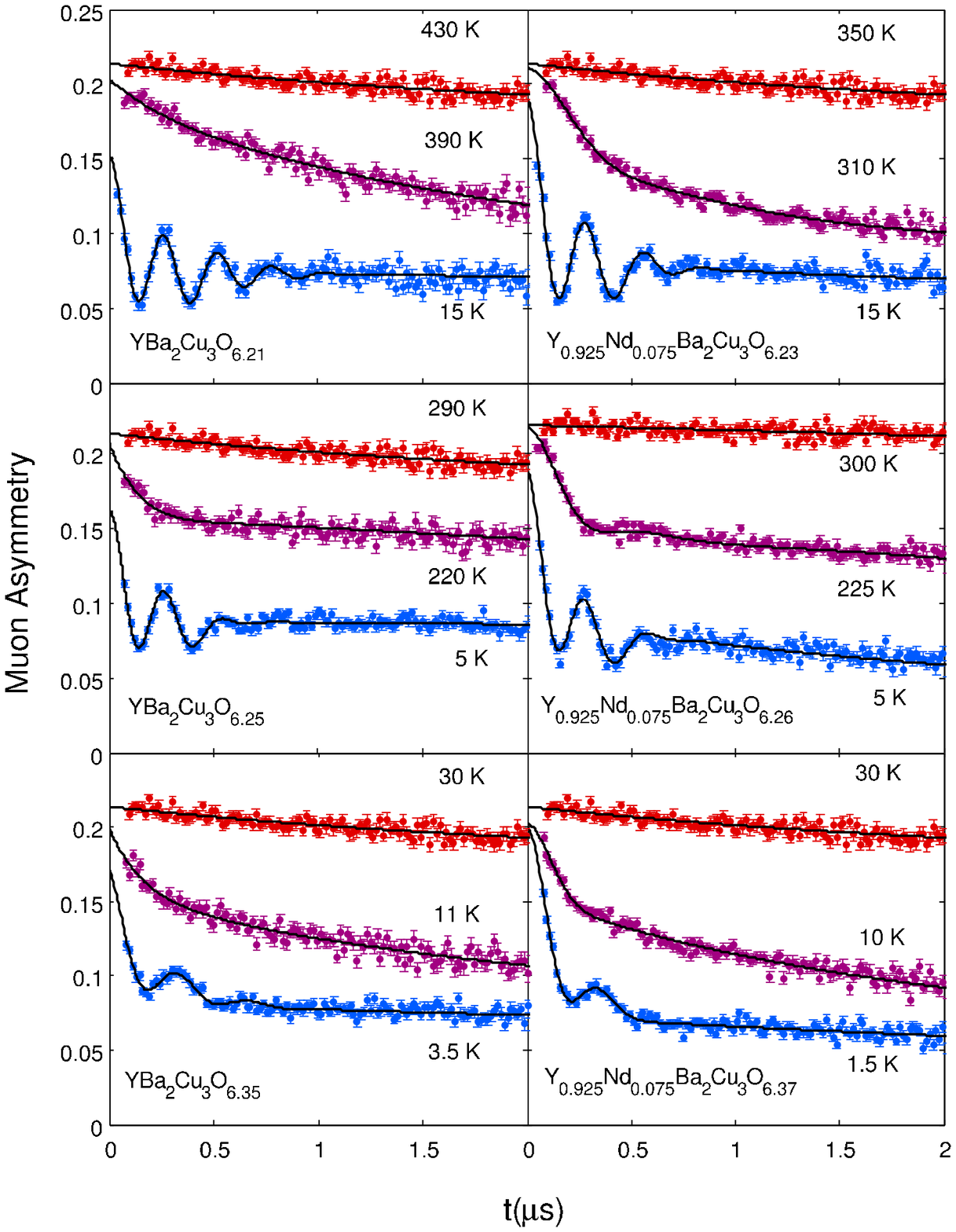}%
\caption{\label{fig:asymmetry} (color on-line) Muon asymmetry above $T_N$, just below and for $T\ll T_N$. Left: three \ybcoy\ samples, right: three corresponding \yndbco\ samples, from nearly undoped (top) to above the onset density for superconductivity.}
\end{figure}

The positron emission in the muon decay is asymmetric, since it is preferentially directed along the muon spin $\bm I$. For a polarized muon ensemble the measured asymmetry is thus directly proportional\cite{Schenck1986} to the average spin self-correlation function,  $A(t)=A_0\overline{I_z(0)I_z(t)}$, where $\hat z$ is the axis of the detector and $A_0$ an experimental parameter subject to calibration.

Typical asymmetry functions in magnetically ordered and paramagnetic phases are shown in Fig.~\ref{fig:asymmetry}. The left column refers to Y100\%,
the right one to Nd7.5\% (very similar data are obtained on Eu8\% as well).
From top to bottom three different samples for each class of compounds span
the magnetic phase diagram, from very low to intermediate doping ($h\approx 0.02$, $h\approx 0.04$, and $h \approx 0.07$ respectively).
In each panel the time evolution of the muon asymmetry is shown for three temperatures, the lowest, one close to the magnetic
transition and one just above it.

The self-correlation function yields a simple relaxing term, if the internal magnetic field $\bm{B}_\mu$ is either vanishing or longitudinal to the initial polarization direction ${\hat I}(0)$, and an additional precessing term, if $\bm{B}_\mu$ is transverse.  Therefore in all six panels the best fit of the sample asymmetry at the lowest temperature, well below the magnetic transition, show both a simple relaxation and precessions, following
\begin{equation}
A(t) = A_0 \left[\sum_i f^t_i e^{-(\sigma_i^2 t^2)/2 } \cos \gamma B_it + f^\ell e^{-{\lambda t}}\right],
\label{eq:lowT}
\end{equation}
where $\gamma=0.8514\cdot 10^9$ (sT)$^{-1}$ is the muon magnetogyric ratio,  $f^{t,\ell}$ are the transverse and longitudinal muon fractions, experiencing relaxation rates $\sigma_i$ and $\lambda$, respectively. The index $i$ spans the two muon sites, AO and CO (Appendix \ref{sec:localfield}), where muons probe two distinct local fields $B_{AO}$ and $B_{CO}$. The longitudinal term does not distinguish the two sites.
The Gaussian form of the transverse relaxation indicates a distribution of static internal fields $p(B_i)$, whose second moment $\left\langle \Delta {B_i}^2 \right\rangle$ is connected to the rate by
$\sigma_i =\gamma  \left\langle \Delta {B_i}^2 \right\rangle^{1/2}$. The experimental data generally include also a small non-relaxing term due to muons stopping outside the sample.

A simple geometric argument shows\cite{Schenck1986} that for a single site in a magnetically ordered polycrystal the experimental muon fractions correspond to an average over all crystal orientations that yields $f^\ell=1/3$ and $f^t=2/3$ (in our case $f^t$ must be replaced by $f^t_{AO}+f^t_{CO}$).
When magnetic and non-magnetic phases are present in the sample, since muons are evenly implanted in both, their fractions reflect the relative volumes of the two phases. The relative magnetic volume  $v_{m}=V_m/V_{tot}$ is given by $v_m=3(1-f^\ell)/2=3f^t/2$, which is the quantity shown in Fig.~\ref{fig:threemagnetizations}b. Its smooth drop across the magnetic transition temperature signals some oxygen inhomogeneity, producing a Gaussian distribution\cite{SannaSSC2003,SannaPRL2004} of mean $T_m$ and width $\Delta T_m$, hence the relative magnetic volume   can be fitted to the function $v_m(T)=[1-\mbox{\rm erf}((T-T_m)/\sqrt2\Delta T_m)]/2$.

The middle plot in each panel of Fig.~\ref{fig:asymmetry} shows data just below $T_m$ characterized by over-damped precessions and increased value of the longitudinal fraction $f^\ell$. Best fits allow only for one undistinguished transverse fraction, with $B=\sqrt{\Delta B^2}$.

At the highest temperatures, well above the magnetic transition, samples are fully in the paramagnetic state and the sample asymmetry is fitted to the
function
\begin{equation}   \label{eq:highT}
A(t) = A_0 e^{-({\sigma_n}^2 t^2)/2 }
\end{equation}
where $\sigma_n$ is the static contribution to the relaxation rate arising from neighboring nuclear moments.

Above $T\approx 250$ K muon diffusion sets in, preventing the direct measure of the transverse precessions. For this reason high temperature points are missing in Figs.~\ref{fig:threemagnetizations}a and \ref{fig:fitTA}. However the N\'eel temperature can still be obtained from the longitudinal fractions,\cite{SannaSSC2003} which are not affected by the incipient muon diffusion, as long as an over-damped transverse fraction can still be distinguished.

\begin{acknowledgments}
This work was carried out at the ISIS muon facility (RAL). We are indebted to Marco Grilli, Pietro Carretta, Giuseppe Allodi and specially to Lara Benfatto and  Jos\'e Lorenzana for very fruitful discussions. We also wish to thank E. Gilioli, T. Besagni and G. Calestani for ancillary experimental support.
Partial support of PRIN-06 project {\em } is acknowledged.

\end{acknowledgments}


\bibliography{PRBiso}

\end{document}